\documentclass[a4paper, twocolumn, floatfix, superscriptaddress]{revtex4-2}
\usepackage{bm,color,graphicx,amsmath,txfonts}
\usepackage[colorlinks, citecolor=blue, linkcolor=blue]{hyperref}
\usepackage{siunitx}
\newcommand{\da}{^\dagger}
\newcommand{\ee}{{\rm e}}
\newcommand{\non}{\nonumber}
\begin{document}
\title{Emergence and enhancement of feedback control induced quantum entanglement}
\author{M. Amazioug} 
\affiliation{LPTHE, Department of Physics, Faculty of Sciences, Ibnou Zohr University, Agadir, Morocco}
\author{D. Dutykh} 
\affiliation{Mathematics Department, Khalifa University of Science and Technology, 127788, Abu Dhabi, United Arab Emirates}
\affiliation{Causal Dynamics Pty Ltd, Perth, Australia}
\author{M. Asjad}
\email{asjad\_qau@yahoo.com}
\affiliation{Mathematics Department, Khalifa University of Science and Technology, 127788, Abu Dhabi, United Arab Emirates}
\begin{abstract}
We present a scheme for controlling quantum correlations by applying feedback to the cavity mode that exits a cavity while interacting with a mechanical oscillator and magnons. In a hybrid cavity magnomechanical system with a movable mirror, the proposed coherent feedback scheme allows for the enhancement of both bipartite and tripartite quantum correlations. Moreover, we demonstrate that the resulting entanglement remains robust with respect to ambient temperatures in the presence of coherent feedback control.
\end{abstract}
\date{\today}
\maketitle 
\section{Introduction}
The phenomenon of entanglement is a fundamental characteristic of quantum mechanics and it is used in many quantum technology applications such as quantum teleportation \cite{CHBennett1993}, superdense coding \cite{CHBennett1992}, telecloning \cite{VScarani2005} and quantum cryptography \cite{AKEkert1991}. In this regard, the field of cavity optomechanics has recently gained a lot of attention as a versatile platform
for variety of quantum information processing applications, particularly realizing entanglement between mechanical and optical dgrees of freedom \cite{Palomaki13, DVitali2007}, ground state cooling of the nanomechanical resonator \cite{Teufel2011, OConnell2010}, generation of two-mode squeezing between microwave and optical fileds \cite{Barzanjeh12,  Abdi12, asjad18} and mechanical degrees of freedom \cite{ockeloen2018,Riedinger_2018}. In quantum meteorology, various achievements have been made, including enhanced precision measurements \cite{HXiong12017, HXiong22017} and gravitational waves detection \cite{CMCaves1980, AAbramovici1992, VBraginsky2002}.

Recently, the magnons, as the quanta of collective spin excitations in yttrium iron garnet ($\rm{ Y_3Fe_5O_{12}}$; YIG), are of paramount importance due to their high spin density, low damping rate and great tunability. A magnon, first introduced by Bloch \cite{Bloch1930} is the collective spin wave excitation carrying quantized energy in a magnetically ordered ground state. Moreover, cavity magnomechanics has attracted considerable attention and provided a robust platform where a ferrimagnetic crystal (\emph{i.e.} YIG sphere) is coupled with a microwave cavity \cite{DLachanceQuirion2019, HYYan2022}, phonons \cite{JLi2018, asjadFund23}, and photons \cite{AOsada2018, AOsada2016, XFZhang2016}.
In cavity magnomechanics, a magnon mode (spin wave) is combined with a vibratory deformation mode of a ferromagnet (or ferrimagnet) by the magnetostrictive force, and a microwave cavity mode by the interaction of magnetic dipoles. The magnetostrictive interaction is a dispersion interaction similar to a radiation pressure for a large ferromagnet, where the frequency of the mechanical mode is very lower than the magnon frequency \cite{ZYFan2022, XZhang2016}. The macroscopic quantum entanglement has been extensively studied by exploiting magnetostrictive interaction in cavity magnomechanical systems \cite{JLi2018}. However, entanglement, as a fragile quantum resource, is prone to being destroyed by the surrounding environment. In addition, the generation of entanglement is often limited by the amplification effect in the unstable regime. Thus, it is crucial to design effective schemes for generating and enhancing entanglement.

Recently, coherent feedback has attracting considerable attention in quantum information processing and quantum optics. Coherent feedback is first discussed in all-optical cases \cite{HMWiseman93a,HMWiseman93b}. In optomechanics consisting of a Fabry-Pérot cavity, coherent feedback is dependent on a light field input that travels through an asymmetric beam splitter to reach the cavity. A part of the output electromagnetic field is partially reflected on the mirror and then feed it back to the cavity via a asymmetric beam splitter. Feedback controle is at the heart of a large field of applications such as to cool a mechanical oscillator \cite{SMancini98, JMCourty01, DVitali0204} and entanglement superconducting qubits\cite{DRiste13}. Implementation of coherent feedback is currently realized in optomechanical systems \cite{MRossi17,JBClark17, NKralj17, MRossi18}, trapped ion \cite{PBushev06}, quantum state transfer \cite{amazioug20} and entanglement \cite{JLi17}. Addationaly, parametric feedback is implemented for cooling and trapping single atoms \cite{MKoch10} and trapped nanospheres \cite{JGieseler12,MGGenoni15}.\\

In this paper presents a scheme to control quantum correlations by deploying feedback on the cavity mode that exits the cavity interacting with magnons and a mechanical oscillator. We investigate that both bipartite and tripartite entanglement can be enhanced by implementing a coherent feedback control in an hybrid cavity magnomechanical system, measured using the logarithmic negativity \cite{VidalWerner, Plenio05, adesso, simon}. Specifically, we show that the resulting entanglement can be controlled and improved even in the presence of thermal noises by adjusting the parameters associated with the feedback loop. 

The paper is organized as follows. In Section \ref{model}, we introduce the model of the hybrid cavity-magnomechanical system under consideration and calculate the quantum Langevin equations along with the covariance matrix for the system. In Section \ref{red}  we discuss the stationary entanglement of the mechanical resonator with the mechanical mode of the YIG sphere ($\mathcal{E}_{\rm db}$), the magnon mode ($\mathcal{E}_{\rm dm}$), and the cavity mode ($\mathcal{E}_{\rm da}$) as functions of various parameters of the system. In Section \ref{red} we study genuine tripartite entanglement. We further study the effects of mechanical thermal noise and parameters associated with feed back on the observation of the tripartite quantum correlation. Section \ref{cn} contains the main concluding remarks. 
\begin{figure}[ht!] 
\centering \includegraphics[width=0.49\textwidth]{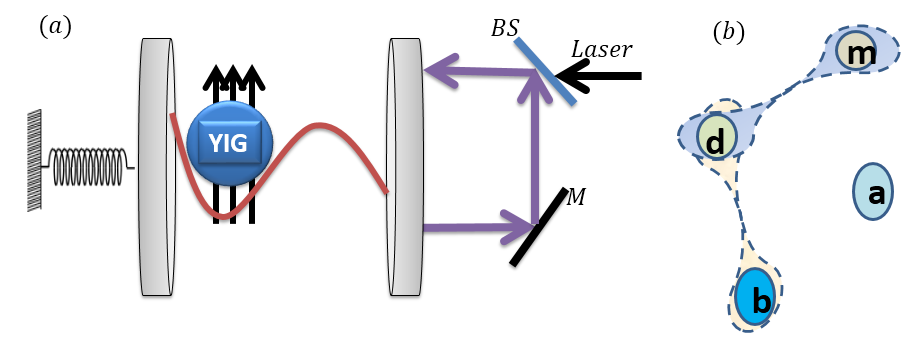}
\caption{(a) Schematics of cavity magnomechanics with movable end mirror via coherent feedback loop. A microwave field (not shown) is implemented to improve magnon-phonon coupling. At YIG sphere site, the magnetic field (along \emph{x}-axis) of the cavity mode, the drive magnetic field (in \emph{y}-direction), and biased magnetic field (\emph{z}-direction) are common perpendicular. In addition, the  coherent feedback loop is achieved through the beam-splitter, where the quantity $t$ and $\tau$ denote the real amplitude transmission and reflection parameters. An input laser field enters in the cavity through an asymmetric beam splitter (BS). The output field is totally reflected on the mirror $M$ and a part of the output field is sent to the cavity via beam splitter.}
\label{fig1}
\end{figure} 
\section{The model}\label{model}
We consider a cavity-magnomechanical (CMM) system consisting of a single ferrimagnetic YIG sphere and a microwave cavity, where the YIG sphere and the microwave cavity are driven separately, as shown in Fig.~\ref{fig1}. The microwave cavity is composed by a fixed mirror (spatially transmitting) and the movable mirror (perfectly reflecting). Due to the bias of the magnetic field along the \emph{y}-axis, YIG sphere excites a large number of magnon modes. At the same time, the magnetization of the YIG sphere causes its deformation, thereby stimulating the mechanical mode of vibration.  The magnons are the quanta of a spin wave in a magnetically ordered state. A magnon mode can be the uniform-precession Kittel mode in the YIG sphere~\cite{XZhang2016,CAPotts2021}. The magnon mode simultaneously couples to a microwave cavity mode via the magnetic-dipole interaction and to a deformation vibration mode via the magnetostrictive-interaction (a radiation pressure-effect-liked). Note that since the wavelength of microwaves is generally much larger than the size of YIG spheres, the radiation pressure coupling of microwave fields to YIG  sphere can be safely ignored.  The Hamiltonian of the CMM system in a rotating frame at the magnon driving frequency $\omega_{0}$ reads as ($\hbar=1$ )
\begin{eqnarray} \label{eq1}
\mathcal{H} &=\Delta_a a^\dagger a + \Delta_m m^\dagger m + \omega_d d^\dagger d + \omega_b b^\dagger b   +  g (a^\dagger m+a m^\dagger) +  
\non\\ 
 & g_{a} a^\dagger a(d + d^\dagger) + g_m m^\dagger m (b+b^\dagger)   + {\rm i} \left( E m^\dagger + \mathcal{E} e^{{\rm i} \phi}  t a^\dagger  -   H.c \right),
 \non\\
 \end{eqnarray}
where $\Delta_{a}:=\omega_{a } - \omega_0$ and  $\Delta_{m}:=\omega_{m } - \omega_0$ are the cavity and magnon detunings. 
The operators $a$, $m$, $b$, and $d$ ($a^\dag$, $m^\dag$, $b^\dag$, and $d^\dag$) are the annihilation (creation) operators related to the cavity, magnon, mechanical mode of the magnon, and  movable end mirror. Here, $g_{\rm a}$ is the coupling of the microwave cavity mode with the movable mirror, $g$ is the coupling of the cavity with the magnonic mode, and $g_{\rm m}$ is the bare magnomechanical coupling which can be significantly improved by directly driving the magnon mode with a microwave field with the coupling strength $E$. The cavity deriving $\mathcal{E}$ is related to the input power $P_L$ by $\mathcal{E}=\sqrt{\frac{2\gamma_a P_L}{\hbar \omega_0}}$, represents the coupling strength between the cavity and the driving field, with $P_L (\omega_0)$ being the power (frequency) of the input laser filed, and $\gamma_a$ the cavity decay rate.  Here $t$ and $\tau$ represent the real amplitude transmission parameters of the beam splitter, and they satisfy the relation $t^2 + \tau^2 = 1$ (with both $t$ and $\tau$ being real and positive). 

In order to describe the complete dynamics of the system we include the dissipation effects. Additionally to the Hamiltonian's description of the system in Eq.(\ref{eq1}), it is subject to noise forces that result from quantum fluctuations of the radiation field and fluctuations of the mechanical and magnon mode's thermal bath. In order to describe the complete dynamics of subsystems involved in this problem, it is an appropriate choice to use the quantum Langevin equations. Therefore, Heisenberg-Langevin equations of motion for optical, mechanical, microwave, and magnon modes are given, respectively, by
\begin{eqnarray}
\dot{a}&=&-(\gamma_{\rm fb}+{\rm i  }\Delta_{\rm fb})a- {\rm i } g m - {\rm i } g_{\rm a} a ( d + d^\dagger)
+\! \!  t\mathcal{E}e^{{\rm i } \phi} +\sqrt{2\gamma_{\rm a} } a_{\rm fb}^{\rm in}, \non \\
\dot{b}&=&-(\gamma_{\rm b}+{\rm i }\omega_{\rm b})b - {\rm i } g_{\rm m} m\da m +\sqrt{2\gamma_b} b_{\rm in},\non\\
\dot{m}&=&-(\gamma_{\rm m}+{\rm i } \Delta_{\rm m}) m -{\rm i } g a - {\rm i } g_{\rm m} (b+b\da)m 
+E + \sqrt{2\gamma_{\rm m}} m_{\rm in},\non \\
\dot{d}&=&-(\gamma_{\rm d}+{\rm i } \omega_{\rm d}) d - {\rm i } g_{\rm a} a^\dagger a +\sqrt{2\gamma_{\rm d}} d_{\rm in}, \label{HL}
\end{eqnarray}
where $\gamma_{\rm \cal{O} }$ (${\cal{O}}=\{d, a, b, m\}$) is the damping rate of the corresponding mode, and ${\cal{O}}_{\rm in}(t)$ denotes the input noise, which is zero-mean $\langle {\rm \cal{O}}_{\rm in}(t)\rangle =0$. For a large value of the mechanical quality factor $Q_{\rm b}=\omega_{\rm b}/\gamma_{\rm b} \gg 1$ and $Q_{\rm d}=\omega_{\rm d}/\gamma_{\rm d} \gg 1$~\cite{DVitali2001}, the mechanical baths can be assumed of Markovian type. Here, $\gamma_{\rm fb}=\gamma_{\rm a}(1-2\tau\cos{\phi})$ and $\Delta_{\rm fb}=\Delta_{\rm a}-2\gamma_{\rm a}\tau\sin{\phi}$ are, respectively, the effective cavity decay rate and the detuning with $\phi$ describes the phase shift generated by the reflectivity of the output field on the mirrors. The output field $a^{\rm out}$ and the cavity field $a$ are related via standard input-output relation $a^{\rm out} = \sqrt{2\gamma_{\rm a} } a - t a^{\rm in}$ \cite{DFWalls1998}. Then, the effective input noise operator $a^{\rm in}_{\rm fb}$, describing the input field induced via the coherent feedback, is given by $a^{\rm in}_{\rm fb} =\tau e^{ {\rm i} \phi}a^{\rm in} + t a^{\rm in}$ which is characterized by the non-zero correlations
\begin{eqnarray}
\langle a_{\rm fb}^{\rm in}(t) \, a_{\rm fb}^{\rm in \dag}(t')\rangle &=& t^2 |1-\tau e^{ {\rm i} \phi}|^2 \left(N_{\rm a}+1\right) \,\delta(t{-}t'),\\
\langle a_{\rm fb}^{\rm in \dag}(t) \, a_{\rm fb}^{\rm in}(t')\rangle \,\,&{=}&\,\, t^2 |1-\tau e^{ {\rm i} \phi}|^2 N_{\rm a} \, \delta(t{-}t'),
\end{eqnarray}
where $N_{\rm a}=\big(\ee^{\hbar \omega_a/k_B T}-1\big)^{-1}$. Here, ${\rm \cal{O}}_{\rm in}(t)$ (${\rm \cal{O}} \in \{d, b, m\}$) is the input nose operator with only nonzero correlations $\langle{\rm \cal{O}}_{\rm in}(t) {\rm \cal{O}}\da_{\rm in}(t')\rangle =(N_{{\rm \cal{O}}}+1)\delta(t-t')$ and  $\langle{\rm \cal{O}}\da_{\rm in}(t) {\rm \cal{O}}_{\rm in}(t')\rangle =N_{{\rm \cal{O}}}\delta(t-t')$, where $N_{{\rm \cal{O}}}=\big( \ee^{\hbar \omega_{k}/k_B T_{\rm \cal{O}}}-1\big)^{-1}$ being the mean thermal occupation number with $k_B$ is the Boltzmann constant and $T_{\rm \cal{O}}$ is the ambient temperature for ${\rm \cal{O}}^{\rm th}$ mode.
\begin{figure*}[ht!] 
\includegraphics[width=2.1 \columnwidth]{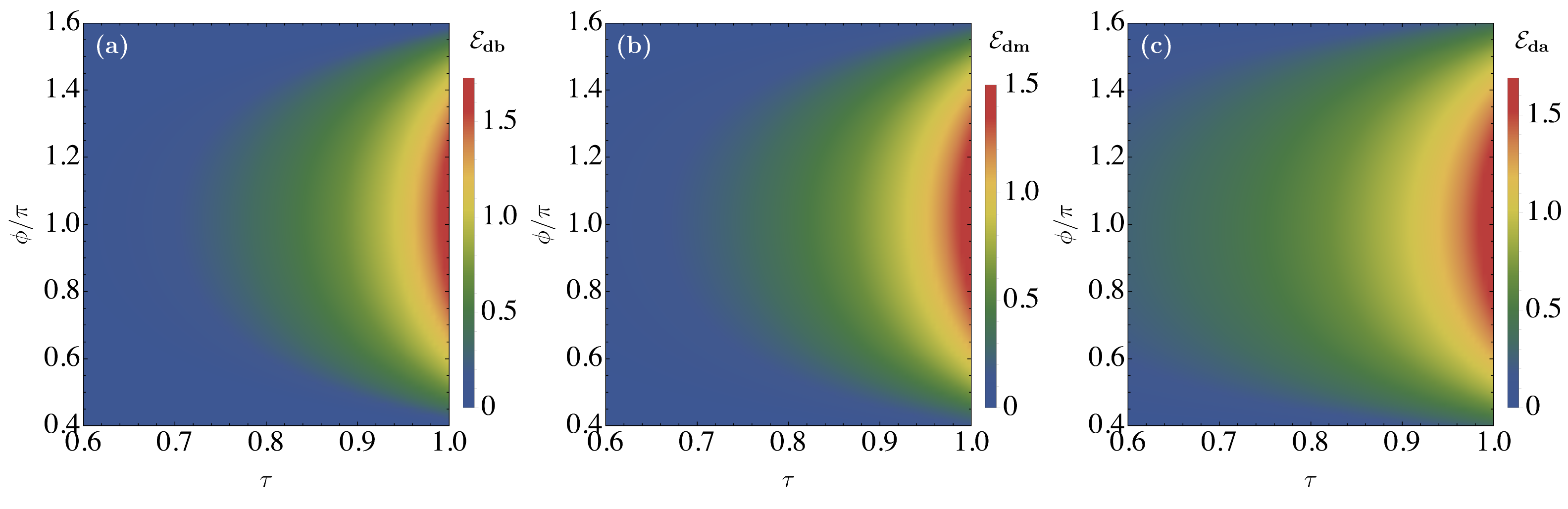}
\caption{(a) Density plot of the Logarithmic negativity $\mathcal{E}_{\rm db}$, (b) $\mathcal{E}_{\rm dm}$ and (c) $\mathcal{E}_{\rm da}$ as a function of reflection coefficient $\tau$ and phase $\phi/\pi$ for a fixed value of detuning $\Delta_{\rm m}=\Delta_{\rm a}=\omega_{\rm b}$.The temperature of the two phononic environments is fixed at $T_{\rm d}=T_{\rm b}=100$ mK.
}
\label{fig2}
\end{figure*}

The nonlinear quantum Langevin equations (QLEs) are in general non-solvable analytically. Using the strong pump for the magnon mode, the QLEs (\ref{HL}) can be linearized by writing each operator $\mathcal{O}$ (${\rm \mathcal{O}} \in \{d,a,b,m \}$) as the sum of its steady-state value and the corresponding small fluctuation operators with zero mean value, \emph{i.e.} $\mathcal{O} \equiv \delta \mathcal{O}+ \ \mathcal{O}_s$. The QLEs (\ref{HL}) are then separated into two sets of equations for classical averages and for quantum fluctuations, respectively. The steady-state averages can be obtained by solving the following equations
\begin{eqnarray}
a_{\rm s} &=& -{\rm i}( g  \, m_{\rm s}  + t \mathcal{E}e^{{\rm i} \phi})/(\gamma_{\rm fb}
+{\rm i} \tilde{\Delta}_{\rm fb}), 
b_{\rm s}=  -{\rm i} g_{\rm m} | m_{\rm s}|^2 /(\gamma_{\rm b}+{\rm i} \omega_{\rm b}), \non \\ 
d_{\rm s}&=&-{\rm i} g_{\rm a} | a_{\rm s}|^2/(\gamma_{\rm d}+{\rm i} \omega_{\rm d}), \quad m_{\rm s} = {(E- {\rm i} g a_{\rm s}) /(\gamma_{\rm m}+{\rm i}\tilde{\Delta}_{\rm m})}, 
\end{eqnarray}
where $\tilde{\Delta}_{\rm fb} = \Delta_{\rm fb} + 2g_{a}{\rm Re} [d_{\rm s}]$ and $\tilde{\Delta}_{\rm m} = \Delta_{\rm m} + 2 g_{\rm m} {\rm Re}\, [b_{\rm s}] $ is the effective detuning by including the magnetostriction. Typically, these frequency shifts are small~\cite{XZhang2016, CAPotts2021,DLachanceQuirion2019}, \emph{i.e.} $|\tilde{\Delta}_{\rm m} - \Delta_{ \rm m}| \ll \Delta_{\rm m} $. Therefore, from now on we can assume $\tilde{\Delta}_{\rm m} \simeq \Delta_{\rm m}$ and $\tilde{\Delta}_{\rm fb} \simeq \Delta_{\rm fb}$.  The explicit expression of the steady-state value $ m_{\rm s} $ is given by
\begin{equation}
m_{\rm s} =  \frac{ E (\gamma_{\rm fb} + {\rm i} \Delta_{\rm fb})}{ g^2  +  \{ \gamma_{\rm m} + {\rm i} \Delta_{\rm m}\} \{ \gamma_{\rm fb} + {\rm i} \Delta_{\rm fb}\} },
\end{equation}
which is generally complex, and its phase can be adjusted by varying the phase of the driving field. Then, the corresponding linearized QLEs describing the dynamics of the quantum fluctuations are given by (for simplicity, we use $\mathcal{O}$ to denote fluctuation operators $\delta \mathcal{O}$)
\begin{eqnarray}  \label{QELs}
\delta\dot{a}&=&-(\gamma_{\rm fb}+{\rm i} \tilde{\Delta}_{\rm fb}) \delta a - {\rm i  }G_{\rm a} (\delta d + \delta d\da) - {\rm i} g  m+\sqrt{2\gamma_{\rm a}}\delta a_{\rm fb}^{ \rm in},\non\\
\delta\dot{b}&=&-(\gamma_{\rm b}+ {\rm i} \omega_{\rm b}) \delta b  - { {\rm i}  (G_{\rm m}^* \delta m+ G_{\rm m} \delta m\da) }+\sqrt{2\gamma_b}\delta b_{\rm in},\non\\
\delta\dot{m}&=&-(\gamma_{\rm m}+{\rm i} \Delta_{\rm m}) \delta m - {\rm i} g  \delta a - {\rm i} G_{\rm m} (\delta b+\delta b^\dagger) +  \sqrt{2\gamma_{\rm m}} \delta m_{\rm in},\non\\
\delta\dot{d}&=&-(\gamma_{\rm d}+{\rm i} \omega_{\rm d}) \delta d - { {\rm i} G_{\rm a} ( \delta a + \delta a \da) }+\sqrt{2\gamma_d}\delta d_{ \rm in}, 
\end{eqnarray}
where $G_{\rm a} =g_{\rm a}  a_{\rm s} $ represents the effective optomechanical coupling strength, and $G_{\rm m} = g_{\rm m} m_{\rm s}$ denotes the effective magnomechanical coupling strength. The QLEs (\ref{QELs}) are linear with respect to annihilation and creation operators. Therefore, it is convenient to introduce the quadrature fluctuations, denoted as $\delta q$ and $\delta p$, for the cavity, magnon and two mechanical modes, which are defined as $\delta{\mathcal{O}}=( \delta q_\mathcal{O} + {\rm i} \delta p_\mathcal{O})/\sqrt{2}$ with ${\rm \mathcal{O}}\in \{a,b,d, m\}$, satisfy the commutation relation $[\delta q_\mathcal{O}, \delta p_\mathcal{O} ]={\rm i}$ (a similar definition applies for corresponding input noise operators $q_\mathcal{O}^{\rm in}$ and $p_\mathcal{O}^{\rm in}$ ). Subsequently, the linearized  QLEs describing the quadrature fluctuations $(\delta q_{\rm d},\delta p_{\rm d},\delta q_{\rm a},\delta p_{\rm a},\delta q_{\rm m},\delta p_{\rm m},\delta q_{\rm b},\delta p_{\rm b})$ can be expressed in a compact matrix form as
\begin{equation}
\dot{\mathcal{R}}(t)=A \mathcal{R}(t)+\mathcal{D} \mathcal{R}_{\rm in} (t),  \label{MatrixForm}
\end{equation}
where $\mathcal{R}(t)=[\delta q_{\rm d}(t),\delta p_{\rm d}(t),\delta q_{\rm a}(t),\delta p_{\rm a}(t),\delta q_{\rm m}(t),\delta p_{\rm m}(t),\delta q_{\rm b}(t),\\ \delta p_{\rm b}(t)]^{\top}$  (where $\top$ represents \emph{transposition})  is the vector of the quadrature fluctuations, $\mathcal{R}_{\rm in}(t)=[q_{\rm d}^{\rm in}(t),p_{\rm d}^{\rm in}(t), q_{a}^{\rm in}(t), p_{a}^{\rm in}(t),q_{\rm m}^{\rm in}(t),p_{\rm m}^{in}(t), q_{\rm b}^{\rm in}(t), p_{\rm b}^{\rm in}(t)]^{\top}$ is the vector for the corresponding input noise quadratureand $\mathcal{D}={\rm diag}[\sqrt{2\gamma_{\rm d}},\sqrt{2\gamma_{\rm d}}, \sqrt{2\gamma_{\rm a}},\sqrt{2\gamma_a} \sqrt{2\gamma_{\rm m}}, \sqrt{2\gamma_{\rm m}},\sqrt{2\gamma_{\rm b}} ,\sqrt{2\gamma_{\rm b}}]$. Here, $A$ is the drift matrix which is given by
\begin{equation} \label{driftmatrix:A}
A=
\begin{pmatrix}
	-\gamma_{\rm d} & \omega_{\rm d} & 0  & 0 & 0 & 0 & 0  & 0\\
    -\omega_{\rm d} & -\gamma_{\rm d} & -G_{\rm a}  & 0 & 0 & 0 & 0  & 0\\
    0 & 0 & -\gamma_{\rm fb}  & \Delta_{\rm fb} & 0 & g & 0  & 0\\
    -G_{a} & 0 & -\Delta_{fb}  & -\gamma_{\rm fb} & -g & 0 & 0  & 0\\
    0 & 0 & 0  & g & -\gamma_{\rm m} & \Delta_{\rm m} & 0  & 0\\
    0 & 0 & -g  & 0 & -\Delta_{\rm m} & -\gamma_{\rm m} & - G_{\rm m}  & 0\\
    0 & 0 & 0  & g & 0 & 0 & -\gamma_{\rm b}  & \omega_b\\
    0 & 0 & 0  & 0 & 0 & G_{\rm m} &  -\omega_{\rm b} & -\gamma_{\rm b}
\end{pmatrix}.
\end{equation} 
The system is stable if the real parts of all the eigenvalues of the drift matrix $A$ are negative. This corresponds to the so-called Routh-Hurwitz criterion \cite{EXDeJesus1987}. 
\begin{figure*}[ht!] 
\includegraphics[width=2.1 \columnwidth]{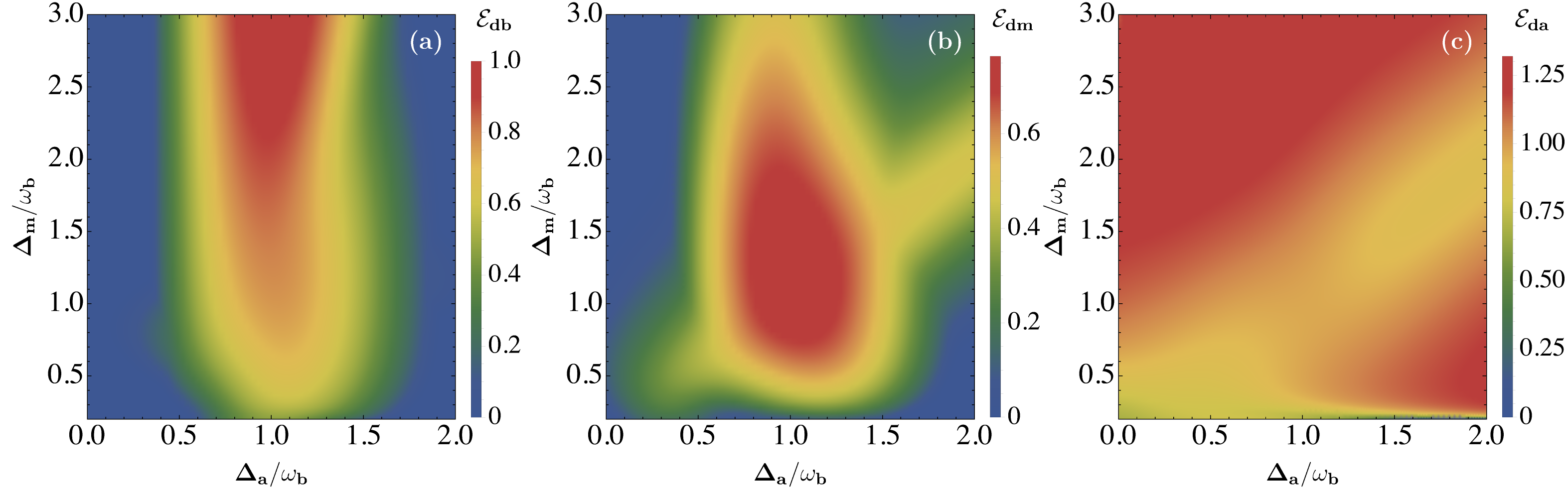}
\caption{Density plot of the three bipartite entanglements (a) entanglement of the moving end mirror with mechanical mode of the YIG sphere $\mathcal{E}_{\rm db}$, (b) magnon mode $\mathcal{E}_{\rm dm}$ and (c) $\mathcal{E}_{\rm da}$ as a function of normalized detunings $\Delta_{\rm m}/\omega_{\rm b}$ and $\Delta_{\rm a}/\omega_{\rm b}$ in the presence of coherent feedback with reflection coefficient $\tau=0.9$ and phase $\phi=\pi$.  All other parameters are the same as in Fig.\ref{fig2}.}
\label{fig3}
\end{figure*}
\section{STEADY STATE ENTANGLEMENT}\label{red}
In this section, we explore three stationary bipartite entanglements: between the mechanical resonator and the mechanical mode of the magnon YIG sphere, the mechanical resonator and the magnon mode, and the mechanical resonator and the cavity mode. Since the quantum noises are white in nature and the dynamics is linearized, the state of the system will be a continuous variable (CV) Gaussian state. Therefore, the steady state of the system is completely characterized by an $8\times 8$ covariance matrix (CM) $\sigma$, defined as $\sigma_{\rm ij}(t)=\langle \mathcal{R}_{\rm i}(t)\mathcal{R}_{\rm j}(t^{\prime })+\mathcal{R}_{\rm j}(t^{\prime })\mathcal{R}_{\rm i}(t) \rangle/2$. Then, the steady state of the $\sigma$ can be obtained by solving directly the Lyapunov equation~\cite{DVitali2007}: 
\begin{equation}
A \sigma+\sigma A^{\top}=-\cal{B}, \label{LyapEq}
\end{equation}
where $\cal{B}$ is the diffusion matrix defined by $\mathcal{B}_{\rm ij}\delta (t-t^{\prime})=\langle \mathcal{R}_{{\rm in}_{\rm i}}(t) \mathcal{R}_{{\rm in}_{\rm i}}(t^{\prime })+\mathcal{R}_{{\rm in}_{\rm i}}(t^{\prime })\mathcal{R}_{{\rm in}_{\rm i}}(t) \rangle/2$, reads as
\begin{eqnarray}\label{eq:D}
\mathcal{B} &=& \text{diag}[\gamma_{\rm d} {\cal N}_{ \rm d},\gamma_{\rm d}  {\cal N}_{ \rm d}, \gamma_{\rm a_{eff}}  {\cal N}_{ \rm a},
\gamma_{\rm a_{\rm eff}} {\cal N}_{ \rm a} , \gamma_{\rm m}  {\cal N}_{ \rm m},\gamma_{\rm m}  {\cal N}_{ \rm m}, \gamma_b {\cal N}_{ \rm b},\nonumber \\
& & \gamma_{\rm b}   {\cal N}_{ \rm b}],
\end{eqnarray} 
where $\gamma_{\rm a_{eff}}=\gamma_{\rm a} t^2[ \{1-\tau \cos(\phi)\}^2+\tau^2 \sin^2(\phi)]$ and ${\cal N}_{\cal O}=2 N_{\cal O}+1$ with $ {\cal O}=$d, a, m and b.
\subsection{Stationary Bipartite Entanglement}
The entanglement between any two of the four bipartite states can be quantified by tracing out two of the four modes and considering the logarithmic negativity, $\mathcal{E}_{\rm N}$ \cite{adesso, simon}. We denote logarithmic negativity for the moving end mirror-mechanical mode of the magnon, magnon-moving end mirror and moving end mirror-cavity mode bimodal parition as $\mathcal{E}_{\rm db}$, $\mathcal{E}_{\rm dm}$ and $\mathcal{E}_{\rm da}$, respectively.  The reduced state is still Gaussian and fully characterized by  its $4 \times 4$ covariance matrix $\Sigma_{\rm i}$ where $\rm i \in \{ \rm db, dm, da\}$.Then the covariance matrix $\Sigma_{\rm i}$ associated with any of the two modes is given by
\begin{equation} \label{eq:Sigmamm}
\Sigma_{\rm i} =
\begin{pmatrix}
	\mathcal{X}_{\rm i} & \mathcal{Z}_{\rm i} \\
    \mathcal{Z}^T_{\rm i} & \mathcal{Y}_{\rm i}  
\end{pmatrix},
\end{equation} 
where $\mathcal{X}_{\rm i}$ and $\mathcal{Y}_{\rm i}$ are $2\times 2$ sub-matrices describing the auto-correlations between the two modes. The $2 \times 2$ sub-matrix $\mathcal{Z}_{\rm i}$ in Eq. (\ref{eq:Sigmamm}) denotes the cross-correlations between the two modes.
The logarithmic negativity $\mathcal{E}_{\rm i}$ is a measure of the entanglement in the bipartite subsystem in the CV system, it can be written as \cite{VidalWerner, Plenio05, adesso, simon}
\begin{equation} \label{eq:37}
	\mathcal{E}_{\rm i}=\max\left(0,-\log(2\nu^-_{\rm i})\right),
\end{equation}
with $\nu^-_{\rm i}$ being the smallest symplectic eigenvalue of the partial transposed covariance matrix $\Sigma_{\rm i}$ of the two modes 
\begin{equation} \label{eq:38}
\nu^-_{\rm i}=2^{\frac{-1}{2}} \left(\Gamma_{\rm i} - \sqrt{\Gamma^2_{\rm i}-4\det\Sigma_{\rm i}}\right)^{\frac{1}{2}},
\end{equation}
where $\Gamma_{\rm i}=\det \mathcal{X}_{\rm i} + \det \mathcal{Y}_{\rm i} - 2\det \mathcal{Z}_{\rm i}$. The two modes are entangled if the condition $\nu^-_{\rm i}<1/2$ (\emph{i.e.} when $\mathcal{E}_{\rm i}>0$) is satisfied. 

We discuss steady state quantum correlations between two and three modes under various effects by taking in consideration experimental values reported in \cite{JLiatom2023}: $\omega_{\rm a}/2\pi= 10$ \unit{GHz}, $\omega_{\rm b}/2\pi= 10 \,\unit{MHz}$, $\gamma_{\rm a}/2\pi=1 \unit{MHz}$, $\gamma_{\rm b}/2\pi=100 \,\unit{Hz}$, $\gamma_{\rm m}/2\pi=1$ MHz, $g/2\pi= 3.2$ MHz, $G_{\rm m}/2\pi= \,4.8 \unit{MHz}$, considering $\omega_{\rm d}/2\pi= 10 \unit{MHz}$, $\gamma_{\rm d}/2\pi=100 \unit{Hz}$ and $G_{\rm a}/2\pi = 3.2$ \unit{MHz} and  $T_{\rm a} = T_{\rm m}=10\, \unit{ m K}$ is the ambient temperature for the magnon and microwave cavity modes which are assumed to be equal.

In Fig. \ref{fig2}  we plot the stationary entanglement between the movable mirror and the mechanical mode of the YIG sphere Fig. \ref{fig2}(a)), the movable mirror and the magnon mode (Fig. \ref{fig2}(b)), and the movable mirror and the photon (Fig.\ref{fig2}(c)). These entanglements are measured by using the logarithmic negativities $\mathcal{E}_{\rm  db}$, $\mathcal{E}_{\rm  dm}$, and $\mathcal{E}_{\rm  da}$, respectively. In Fig. \ref{fig2}, we plot $\mathcal{E}_{\rm  db}$ as a function of the reflection coefficient $\tau$ and phase $\phi/\pi$ for fixed values of the detunings $\Delta_{\rm a}=\Delta_{\rm a}=\omega_{\rm b}$. The entanglement measure between movable end mirror and mechanical mode of the YIG sphere increases with $\tau$ and reaches its maximum value for $\phi \approx \pi$ and reflection coefficient $\tau \approx 1.0$. Similarly, the stationary entanglement of the movable mirror with the magnon and cavity modes is measured using the logarithmic negativity $\mathcal{E}_{\rm dm}$ and $\mathcal{E}_{\rm  da}$ as a functions of $\tau$ and $\phi/\pi$, as shown in Figs. \ref{fig2}(b) and \ref{fig2}(c), respectively. The features observed in the entanglement of the movable mirror with the magnon and cavity modes are similar to those present in $\mathcal{E}_{\rm  db}$. It is evident from Fig. \ref{fig2} that the entanglement between different bimodes can be enhanced with a suitable choice of phase $\phi$ and reflection coefficient $\tau$.  The three bipartite entanglements reach their maximum value when $\gamma_{\rm fb}=\gamma_{\rm a}+2 \tau\gamma_{\rm a}$ and $\Delta_{\rm fb}=\Delta_{\rm a}$, \emph{i.e.} when $\phi=\pi$.

In the presence of coherent feedback, the entanglement of the moving end mirror with mechanical mode of the YIG sphere $\mathcal{E}_{\rm db}$, magnon $\mathcal{E}_{\rm db}$ and cavity $\mathcal{E}_{\rm db}$ modes as a function of normalized detunings $\Delta_{\rm m}/\omega_{\rm b}$ and $\Delta_{\rm a}/\omega_{\rm b}$ are shown Fig. \ref{fig3}. It is clear from Fig. \ref{fig3}(a), Fig. \ref{fig3}(b), and Fig. \ref{fig3}(c) that the three bipartite entanglements, $\mathcal{E}_{\rm db}$, $\mathcal{E}_{\rm dm}$, and $\mathcal{E}_{\rm da}$, in the presence of coherent feedback exist only within a small interval of values for $\Delta_{\rm a}$, with the maximum occurring around $\Delta_{\rm a} \approx \omega_{\rm b}$.

\begin{figure}[ht!] 
\includegraphics[width=0.95\columnwidth]{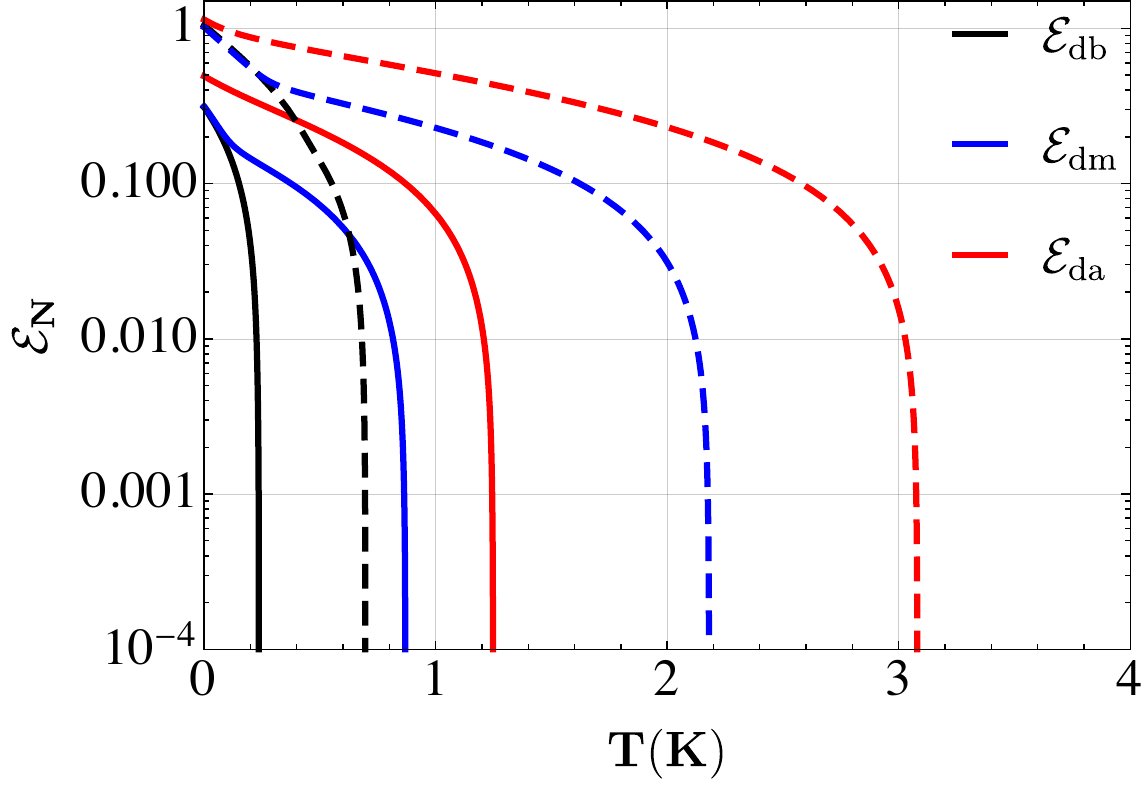}
\caption{Plot of three bipartite entanglements: $\mathcal{E}_{\rm db}$ (black curves), $\mathcal{E}_{\rm dm}$ (blue curves), and $\mathcal{E}_{\rm da}$ (red curves) as functions of the temperature T (K) for different values of the reflection coefficient $\tau=0.7$ (solid curves) and $\tau=0.9$ (dashed curves), with a fixed value of phase $\phi =\pi$ and detuning $\Delta_{\rm a}=\Delta_{\rm a}=\omega_{\rm b}$. All the others parameters are the same as in Fig.\ref{fig2}.
}
\label{fig4}
\end{figure} 

It is important to study the robustness of the entanglement with respect to the environmental temperature $T$, \emph{i.e.} $T = T_{\rm b} = T_{\rm d}$.  In Fig.\ref{fig4} we plot three bipartite entanglements: $\mathcal{E}_{\rm db}$ (black curves), $\mathcal{E}_{\rm dm}$ (blue curves), and $\mathcal{E}_{\rm da}$ (red curves) against the environmental temperature $T$ for $\tau=0.7$ (solid curves) and $\tau=0.9$ (dashed curves) when $\Delta_{\rm a}=\Delta_{\rm m}=\omega_{\rm b}$ and the phase $\phi=\pi$. For  $\tau=0.7$, a significant amount of the entanglement between the moving end mirror and the mechanical mode of the YIG sphere, moving end mirror and magnon mode, and moving end mirror with cavity mode are present up to temperatures $T$ (approximately) of $0.3\, \unit{K}$, $0.9\, \unit{K}$, and $1.2\,\unit{K}$, as shown in Fig. \ref{fig4} (solid curves). However, the three bipartite entanglements $\mathcal{E}_{\rm db}$, $\mathcal{E}_{\rm dm}$, and $\mathcal{E}_{\rm da}$ are more robust against the environmental temperature $T$ and persist up to $0.7\, \unit{K}$ , $2.2 \, \unit{K}$, and $3.1\,  \unit{K}$  when the reflection coefficient is fixed at $\tau=0.9$, as shown in Fig. \ref{fig4} (dashed curves).  
\begin{figure}[ht!] 
\includegraphics[width=0.48\textwidth]{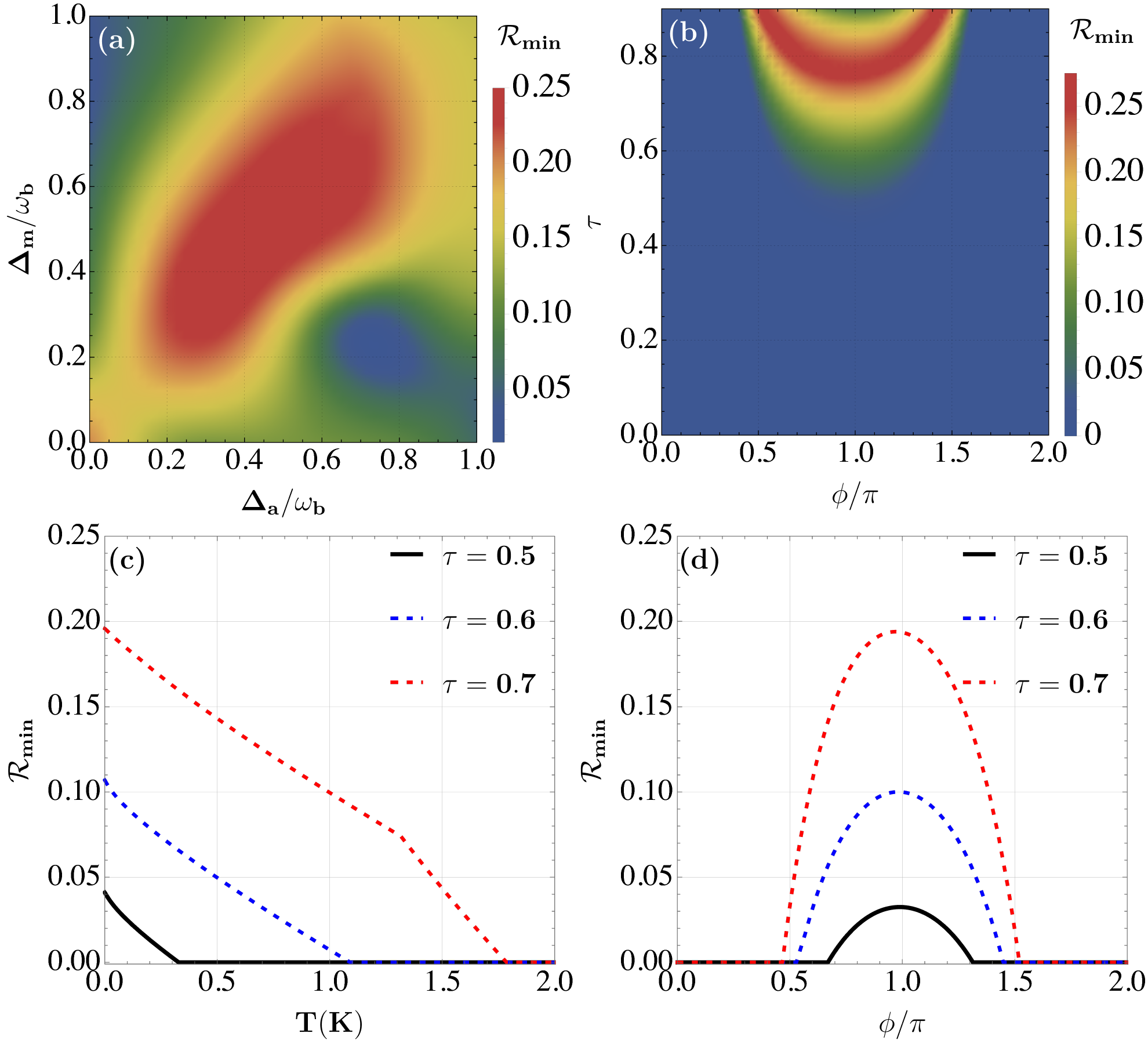}
\caption{(a) Residual contangle $\mathcal{R}_{\rm min}$ as a function of normalized magnon detuning $\Delta_{\rm a}/\omega_{\rm b}$ and cavity detuning $\Delta_{\rm m}/\omega_{\rm b}$ for the fixed values of the phase  $\phi=1.0 \pi$ and the reflection coefficient $\tau=0.7$. (b)  $\mathcal{R}_{\rm min}$ as a function of reflection coefficient $\tau$ and phase $\phi$ for a constant value of the cavity and the magnon detunings $\Delta_{\rm a}=\Delta_{\rm m}=0.5\omega_{\rm b}$. (c) $\mathcal{R}_{\rm min}$ as a function of the environmental temperature T(K), \emph{i.e}, $ {\rm T}={\rm T}_{\rm b}={\rm T}_{\rm d}$ for $\tau=0.5$ (black curve), $\tau=0.6$ (blue dashed curve) and $\tau=0.7$ (red dashed curve) with $\Delta_{\rm a}=\Delta_{\rm m}=0.5\,\omega_{\rm b}$ and $\phi=1.0 \,\pi$. (d) $\mathcal{R}_{\rm min}$ as function of phase $\phi/\pi$ for $\tau=0.5$ (black curve), $\tau=0.6$ (blue dashed curve) and $\tau=0.7$ (red dashed curve) with $\Delta_{\rm a}=\Delta_{\rm m}=0.5\,\omega_{\rm b}$ at fixed value of $ {\rm T}= 100 \, \unit{m K}$ and $\Delta_{\rm a}=\Delta_{\rm m}=0.5\omega_{\rm b}$. All the others parameters are the same as in Fig.\ref{fig2}.
 }
\label{fig5}
\end{figure} 
\subsection{Tripartite Entanglement} 
In this section we discuss the genuine tripartite entanglement and use the residual contangle to examine tripartite movable mirror-magnon-mechanical mode of YIG sphere entanglement \cite{GAdesso2007}, defined as
\begin{equation}
\mathbb{R}^{\rm i|\rm jk} \equiv {\cal E}_{\rm i|\rm jk} - {\cal E}_{\rm i|\rm j} - {\cal E}_{\rm i|\rm k},\quad {\rm i, j, k \in \{d, m, b\}}
\end{equation}
where ${\cal E}_{x|y}$ is the contangle of two modes $x$ and $y$ ($y$ composed of one or two modes), and is quantified as the squared logarithmic negativity. For Gaussian states, the minimum residual contangle offers an exact measurement of tripartite entanglement when $\mathbb{R}^{\rm i|\rm jk}>0$ 
\begin{equation}\label{eqr}
{\cal R}_{\rm min} \equiv {\rm min} \left( \mathbb{R}^{\rm d|\rm mb}, \, \mathbb{R}^{\rm m|\rm db}, \, \mathbb{R}^{\rm b|\rm dm}  \right).
\end{equation}
The quantum entanglement's monogamy is satisfied by the residual contangle, is similar to the Coffman-Kundu-Wootters monogamy inequality~\cite{WKWootters2000} holding for the system of three qubits.  Hence, a \textit{genuine} tripartite entanglement emerges only when the minimum residual entanglement of the system ${\cal R}_{\rm min}$ is greater then \emph{zero},\emph{i.e}. ${\cal R}_{\rm min}\,{>}\,0$.

In Fig.\ref{fig5}, we investigate the minimum value of the residual contangle $\mathcal{R}_{\rm min}$ given in Eq.(\ref{eqr}), concerning various parameters in the system. Fig.\ref{fig5}(a) shows dependence of the contangle $\mathcal{R}_{\rm min}$ on the normalized detunings $\Delta_{\rm  a}/\omega_{\rm b}$ and $\Delta_{\rm m}/\omega_{\rm  b}$ for $\tau=0.75$ and phase $\phi=\pi$. It is evident from Fig. \ref{fig5}(a) that $\mathcal{R}_{\rm min}$ exists only within a limited range of values for $\Delta_{\rm  a}$ and $\Delta_{\rm  m}$, centered around $\Delta_{\rm  m} = \Delta_{\rm  a} \approx 0.5\omega_{\rm  b}$.
Fig.\ref{fig5}(b), shows the dependence of $\mathcal{R}_{\rm min}$ on feedback parameters reflection coefficient $\tau$ and $\phi/\pi$ for fixed values of the detunings  $\Delta_{\rm  a}= \Delta_{\rm  m} =0.5 \omega_{\rm b}$. It is clear that the optimal values for the reflection coefficient and phase for the contangle $\mathcal{R}_{\rm min}$ is correspond to red region in Fig.\ref{fig5}(b) for $\Delta_{\rm  a}= \Delta_{\rm  m} =0.5 \omega_{\rm b}$.

We further study the effects of mechanical thermal noise on the steady state residual contangle $\mathcal{R}_{\rm min}$ as a function of different parameters of the system. In Fig.\ref{fig5}(c), we plot the $\mathcal{R}_{\rm min}$ as a function temperature T (\unit{K}) of the two phononic modes, \emph{i.e.} $T$=$T_{\rm b}$=$T_{\rm d}$ for different values of the reflection coefficient $\tau=0.5$ (black curve), $0.6$ (blue curve) and $0.7$ (red curve) at fixed values of the phase $\phi=0.65 \pi$ and of the detuning $\Delta_{\rm  a}= \Delta_{\rm m} =0.5 \omega_{\rm  b}$. It is noted that $\mathcal{R}_{\rm min}$ is very sensitive to the temperature of mechanical modes and quickly decays for large values of the temperature. In high temperatures, thermal fluctuations always suppress the residual contangle.  In Fig.\ref{fig5}(d), we plot $\mathcal{R}_{\rm min}$ as a function of phase $\phi/\pi$ for different values of reflection coefficient $\tau=0.5$ (black curve), $0.6$ (blue curve) and $0.7$ (red curve) when detuning is fixed at $\Delta_{\rm a}= \Delta_{\rm m} =0.5 \omega_{\rm b}$.  One can easily observe that the stationary residual contangle $\mathcal{R}_{\rm min}$ reaches it maximum value at phase $\phi \approx \,\pi$.
\section{Conclusions}\label{cn}
We have introduced a theoretical scheme for generating stationary bipartite and \textit{genuine} tripartite entanglement in cavity magnomechanical systems with a moving end mirror through coherent feedback. Our study illustrates that the resultant bipartite and \textit{genuine} tripartite quantum correlations can be controlled and significantly improved through the implementation of coherent feedback. Moreover, we have demonstrated the robustness of the resulting quantum entanglement in the presence of environmental temperature associated with phononic modes. We have investigated that the cavity magnomechanical systems could provide a promising platform for the study of macroscopic quantum phenomena just by implementing the appropriate feedback control. The proposed scheme to enhance quantum correlations through coherent feedback can be of interest for applications in quantum information processing. 
\section*{Acknowledgements}
M. Asjad and D. Dutykh have been supported by the Khalifa University of Science and Technology under Award No. FSU-2023-014).
%
\end{document}